\documentclass[reprint,superscriptaddress,amsmath,amssymb,aps,pra,floatfix]{revtex4-2}

\usepackage{graphicx}
\usepackage{dcolumn}
\usepackage{bm}
\usepackage{dsfont}
\usepackage{orcidlink}
\usepackage{mathtools}
\usepackage[capitalise]{cleveref}
\usepackage{xcolor}

\newcommand{\average}[1]{\left\langle#1\right\rangle}
\newcommand{\ket}[1]{\left\vert#1\right\rangle}
\newcommand{\bra}[1]{\left\langle#1\right\vert}
\newcommand{\comm}[2]{\left[#1,#2\right]}
\newcommand{\pTrace}[2]{\underset{#1}{\mathrm{Tr}}\left[#2\right]}
\newcommand{\HH}{\hat{H}}
\newcommand{\Abs}[1]{\left|#1\right|^{2}}

\newcommand{\rperp}{r_{\perp}}
\newcommand{\tstar}{t_{\star}}
\newcommand{\Phin}{\Phi_{n}}

\begin{document}

\preprint{}

\title{Inferring the Number of Quantum State Transfers with Neural Networks}

\author{Hayden Zammit\,\orcidlink{0009-0003-9917-929X}}
\email{hayden.zammit@um.edu.mt}
\affiliation{Department of Physics, University of Malta, Msida MSD 2080, Malta}

\author{Tony J. G. Apollaro\,\orcidlink{0000-0002-9324-9336}}
\affiliation{Department of Physics, University of Malta, Msida MSD 2080, Malta}

\author{Johann A. Briffa\,\orcidlink{0000-0003-3532-9474}}
\affiliation{Department of Communications and Computer Engineering, University of Malta, Msida MSD 2080, Malta}

\author{Gianluca Valentino\,\orcidlink{0000-0003-3864-7785}}
\affiliation{Department of Communications and Computer Engineering, University of Malta, Msida MSD 2080, Malta}

\date{\today}

\begin{abstract}
Repeated-use quantum state transfer (QST) goes beyond the single-use
paradigm by accounting for the memory that accumulates inside the
transmission channel when it is not reset between uses. Consequently,
the map governing the $n^{\mathrm{th}}$ use of the
channel differs from use to use and carries a signature of each previous use. We ask whether this use number can be
recovered from transmission data alone, and show that it can: a simple
feed-forward neural network, given as input the Bloch vectors sent and received over a single use of an $XX$ spin-chain channel,
is able to identify the use number with an accuracy exceeding $90\%$. Exploiting the $U(1)$ symmetry of the channel, we find that a
reduced two-component representation of the transmitted states improves
the accuracy further, to about $95\%$. We characterise the dependence of the classifier on
the size and composition of the training data, finding that the number of sampled states and
of sampled timings are equally important. We also identify the data
set sizes required to reach a given accuracy. Finally, we quantify the severity of the residual misclassifications with an earth-mover distance on the confusion matrix.
\end{abstract}

\maketitle

\section{Introduction}\label{sec_intro}
Quantum state transfer (QST) refers to the task of moving an unknown
quantum state between distant nodes of a quantum network. The use of spin-$\frac{1}{2}$ chains as the connecting medium was
proposed by \citet{Bose2003}, who showed that the free dynamics of an
unmodulated chain is by itself sufficient to carry the quantum information encoded in a qubit from one end
to the other. The appeal of the scheme lies in its economy, since no
external control need be applied during the transfer itself. Subsequent
work established chains whose couplings are engineered so as to achieve
perfect state transfer (PST), for which the transfer fidelity reaches
unity at a specific readout time
\cite{Christandl2004, Kay2006, vinetHowConstructSpin2012}. Experimental
realisations have since been reported in photonic and superconducting
architectures \cite{chapmanExperimentalPerfectState2016a, Li2018a,
zhouQuantumStateTransfer2024}, and the subject remains an active one
(see, for example, \cite{Bose01012007, Nikolopoulos2014} for reviews).
 
The overwhelming majority of these works treat a single use of the
channel. In the standard picture the chain is initialised, one qubit is
transferred, and the chain is then reset before the next transfer is
attempted. Resetting is, however, a time- and resource-consuming
operation \cite{navascuesResettingUncontrolledQuantum2018,
bassmanoftelieDynamicCoolingContemporary2024}. If the chain is instead
used continuously, the residual excitations left behind by earlier
transfers remain within the medium and interfere with later transmissions.
Physically, the channel thereby acquires a memory of its own history, in
the sense that the completely positive map governing the
$n^{\mathrm{th}}$ transfer depends on $n$. Memory effects of this kind
are known to arise in spin chains and in quantum channels more generally
\cite{bayatMemoryEffectsSpinchain2008a, carusoQuantumChannelsMemory2014b,
Giovannetti_2005, benentiEnhancementTransmissionRates2009}, and were
derived analytically for $U(1)$-symmetric channels in
Ref.~\cite{ZammitMemory2026}. In particular, it was shown there that even
small errors in the readout timing, an imperfection which is difficult to
avoid in practice \cite{xuerebImpactImperfectTimekeeping2023,
meierFundamentalAccuracyresolutionTradeoff2023,
robledoHighfidelityProjectiveReadout2011a}, give rise to memory which can degrade the fidelity of subsequent uses considerably.
 
The existence of this memory admits a natural corollary. If each use is
governed by a genuinely different map, then the number of past uses ought
to leave a signature on the transmitted data. In this work we investigate whether this is the case. Given a description of the input and output of a particular transmission, is it possible to determine the number of times the chain has been used, and thereby characterise the memory inside the channel?

The ability to do so has practical consequences. The usage history
of a quantum bus is metadata which one might expect to be buried in the
uncontrolled degrees of freedom of the channel. Recovering it from
transmission data alone would render it accessible without any additional
probing of the medium, and without interrupting the communication task
itself. In particular, such a capability appears relevant to the
calibration and drift-monitoring of a channel, to resource accounting and
scheduling within a quantum network, and, more broadly, to the diagnosis
of a device from its input-output behaviour
\cite{sarovarDetectingCrosstalkErrors2020}.
 
No obvious closed-form estimator exists for this problem. The map depends
on the use number, but it depends also on the readout timing offset, a parameter which is typically not exactly known to the observer.
Furthermore, different combinations of the two can, in principle, produce similar
input-output behaviour; for example, the output at the third use for one timing offset may resemble that at the fourth use for a different timing offset. Problems of this type are naturally addressed by
supervised learning, which has by now been applied widely across the
physical sciences, and within quantum information
\cite{Carleo2019, Dunjko2018}.
 
Neural networks have been used to
perform quantum process tomography, where the exponential cost of the
conventional procedure is ameliorated by a learned representation of the
channel \cite{torlaiQuantumProcessTomography2023}. Learning techniques
have likewise been applied to the characterisation of noise, such as
classifying the non-Markovian noise which disturbs a quantum system
\cite{martinaMachineLearningClassification2023}, and to approximating the
reduced channel which describes the dynamics of a subsystem on
present-day hardware \cite{ceminMachineLearningReduced2024}. The
discrimination of quantum channels has also been framed as a learning
problem, by training variational quantum circuits to
distinguish between a pair of candidate channels
\cite{kardashinQuantumMachineLearning2022}. Feed-forward neural networks have also been used to recover the originally pure states which have undergone noisy evolution \cite{morgilloQuantumStateReconstruction2024}.
 
The problem solved in this work is stated concretely as follows. A PST chain is used
repeatedly, up to five times, and is not reset between uses.
Each use is read out at the same time, offset from the optimal transfer
time by a fixed fractional error of at most $20\%$, so as to model a
timer carrying a constant systematic bias. The state presented to the
chain at each use is a pure state drawn at random from the Haar measure
\cite{Mezzadri2007}. A single labelled example then consists of the Bloch
vector of the sender (the channel input), the Bloch vector of the corresponding receiver (the channel output), and
the number of the use which produced them. The task set to the classifier
is to recover that number from the pair of input-output Bloch vectors. 
 
The main findings are as follows: i) using as input the sender and
receiver Bloch vectors, a feed-forward neural network is able to classify
the use number of a repeatedly used PST chain with an accuracy exceeding
$90\%$, and generalises well to both states and times not seen during
training; ii) for the same data set, network architecture, and number of training epochs, a reduced two-component parameterisation of the sender and
receiver Bloch vector inputs, motivated by the $U(1)$ symmetry of the
channel, improves the accuracy relative to the raw Bloch vector
parameterisation, achieving an overall accuracy of $95\%$; iii) the accuracy depends systematically
on the number of sampled states and on the number of sampled timings, but,
to a very large degree, only through their product, so that the two
contribute equally to the final achievable accuracy; the data set size required for a target
accuracy can therefore be stated as a single number, and the reduced
parameterisation requires between four and seven times fewer examples than
the raw one; and iv) an earth-mover distance evaluated on the confusion
matrix shows that about two thirds of the residual misclassifications fall
on neighbouring uses.
 
The generalisation to unseen states is expected of any adequately
trained classifier. The generalisation to unseen times
is less obvious, since, as previously stated, different numbers of uses and timing offsets can have similar effects on state transfer. That said, the use number is
recoverable at timings never presented during training.
 
The rest of this work is organised as follows. In \cref{sec_model} the
repeated-use QST protocol is reviewed, the family of channels to be
discriminated is derived, and the reduced coordinates suggested by the
$U(1)$ symmetry are introduced. \Cref{sec_methods} states the
classification problem and details the generation of the data, the
architecture of the network, and the metrics against which it is
assessed. The results are presented in \cref{sec_results}, and
\cref{sec_conclusion} concludes.

\section{Repeated-use channel}\label{sec_model}

In this section we define the model of repeated-use QST through a spin-$\frac{1}{2}$ chain which is not reset in between uses. We will first derive the dynamical map for single-use QST. 

\subsection{Single-use QST}\label{sec_single}

We consider QST through a one-dimensional spin-$\frac{1}{2}$ chain of length
$N$ evolving under some $U(1)$-symmetric Hamiltonian $\HH$~\cite{Bose2003}, where $U(1)$ symmetry is equivalent to the condition of the conservation of the total magnetisation,
\begin{equation}\label{eq_u1}
    \comm{\HH}{\textstyle\sum_{i=1}^{N}\hat{\sigma}^{z}_{i}} = 0 ~,
\end{equation}
so that $\HH$ preserves the number of excitations and decomposes
into a direct sum over fixed-excitation sectors. The basic QST task is to transfer the information of a pure state
$\ket{\psi}_{1}=\alpha\ket{0}+\beta\ket{1}$, encoded on the first site, through the chain initialised in $\ket{0\cdots 0}_{2,\dots,N-1}$, to the last site $\ket{0}_{N}$, by allowing the total system to evolve under $\HH$ for a time $t$. The transfer time $t$ is chosen to maximise the similarity between the initialised first-site state $\ket{\psi}_{1}$ and the state of the final site at the time $t$, $\hat{\rho}_{N}(t)$. The standard metric of similarity is the Uhlmann--Jozsa fidelity \cite{Jozsa:1994qja}
\begin{equation}\label{eq. Ulhmann - Josza}
    F(t) = \bra{\psi}_{1}\hat{\rho}_{N}(t)\ket{\psi}_{1} ,
\end{equation} 
whose maximal value of $1$ can only be achieved with PST. For a $U(1)$-symmetric Hamiltonian the dynamics stays inside the zero- and single-excitation sectors. Then the relevant quantities which determine $\hat{\rho}_{N}(t)$ are the transition amplitudes
\begin{equation}\label{eq_amp}
    f_{i}^{j}(t) = \bra{j}e^{-i\HH t}\ket{i}
    = \sum_{k=1}^{N}\nu_{kj}\,\nu_{ki}^{\ast}\,e^{-iE_{k}t}~,
\end{equation}
where $\ket{j}$ denotes the state with a single excitation on site $j$, $E_{k}$ are the single-particle eigenenergies, and $\nu_{ki}$ are the
components of the corresponding eigenstates. In terms of these amplitudes, the state of the last site reads
\begin{align}\label{eq_final_evolved_reciever}
    &\hat{\rho}_{N}(t)=\pTrace{1,2,\dots,N-1}{e^{-i\HH t}\ket{\psi(0)}\bra{\psi(0)}e^{i\HH t}}\nonumber\\
    &= \begin{pmatrix}
    1-|\beta|^{2}|f_{1}^{N}(t)|^{2} & \alpha\beta^{*} f_{1}^{N}(t)^{*} \\
    \alpha^{*}\beta f_{1}^{N}(t) & |\beta|^{2}|f_{1}^{N}(t)|^{2}
    \end{pmatrix}~,
\end{align}
where $\ket{\psi(0)}=\ket{\psi}_{1}\ket{0\cdots 0}_{2,\dots,N}$ is the initial state of the whole chain. This can be written as
\begin{equation}\label{eq_dyn_map_1}
    \begin{pmatrix}
    \rho_{00}\\
    \rho_{01}\\
    \rho_{10}\\
    \rho_{11}
    \end{pmatrix}_{\!N} =
    \begin{pmatrix}
        1 & 0 & 0 & 1-|f|^{2}\\
        0 & f^{*} & 0 & 0\\
        0 & 0 & f & 0\\
        0 & 0 & 0 & |f|^{2}
    \end{pmatrix}\!\!
    \begin{pmatrix}
        |\alpha|^{2}\\ 
        \alpha\beta^{*}\\
        \alpha^{*}\beta\\
        |\beta|^{2}
    \end{pmatrix}_{\!1} ,
\end{equation}
with $f\equiv f_{1}^{N}(t)$, in terms of the completely positive and trace-preserving dynamical map $\Phi$~\cite{Society2004,Lorenzo2022} which satisfies $\hat{\rho}_{N}(t) = \Phi(\hat{\rho}_{1})$. Finally, whereas \cref{eq. Ulhmann - Josza} describes the quality of transferring the specific state prepared on the initial site, it can be extended to measure the overall quality of the transfer scheme itself by evaluating the average over all pure state inputs. For $U(1)$-symmetric chains this takes the closed form~\cite{Bose2003}
\begin{equation}\label{eq_fid1}
    \average{F(t)} = \frac12 + \frac13\,\big|f_{1}^{N}(t)\big|\cos\gamma
    + \frac16\,\big|f_{1}^{N}(t)\big|^{2}~,
\end{equation}
with $\gamma=\arg\{f_{1}^{N}(t)\}$ denoting a phase that can be removed by a fixed local rotation on site $N$.

\subsection{Repeated-use QST}\label{sec_repeated}
\begin{figure}[t]
    \centering
    \includegraphics[width=\linewidth]{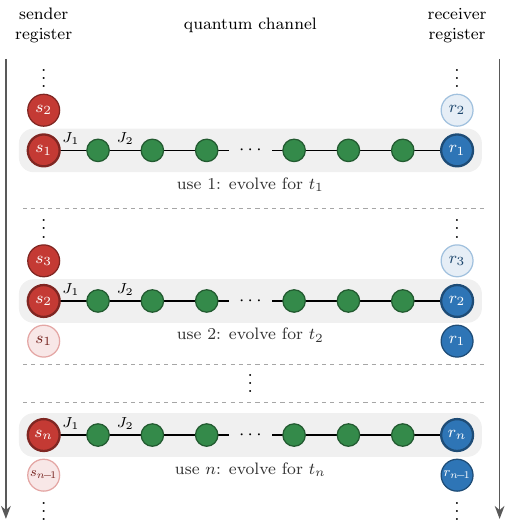}
    \caption{The repeated-use QST protocol. Quantum information held in an $n$-qubit sender register (red) is transferred, one qubit per use, to an $n$-qubit receiver register (blue) through a shared spin-chain channel (green) that evolves for a time $t_i$ at the $i^{\mathrm{th}}$ use and is not reset between uses.}
    \label{fig_model}
\end{figure}

We now use the same physical chain repeatedly, without resetting
the intermediate sites between transfers, following the protocol of
Ref.~\cite{ZammitMemory2026}, as illustrated in \cref{fig_model}. A
register of sender qubits is transferred, one per use, to a register of
receiver qubits: at each use a fresh sender qubit and a fresh
receiver qubit (the latter always in $\ket{0}$) are coupled to the channel, the whole
system evolves for a time $t_{n}$, and the receiver is read out before
the next sender and receiver qubits are attached.

After $n$ uses the state of the channel depends on all of the transfer
times $t_{n;1}=(t_1,\dots,t_n)$ and on the states of all previous senders. However, the states sent at previous uses are, in general, not known. In this scenario it is appropriate to consider the average effect of each imperfect use. Performing this average on the channel state defines the average 
$n^{\mathrm{th}}$-use channel state, which encodes
the best description of the channel available when only the number of past
uses is known. Restricting to Hamiltonians that
are $U(1)$-symmetric and quadratic in fermions---so that all
multi-particle amplitudes factor into Slater determinants of the
single-particle amplitudes~\cite{apollaroQuantumTransferInteracting2022}---the map that sends the sender
at the start of the $n^{\mathrm{th}}$ use to the receiver at its end can be
written in the closed form
\begin{widetext}
\begin{equation}\label{eq_mapn}
\Phin(t_{n;1})=
\begin{pmatrix}
 1-\tfrac12\sum_{i=1}^{n-1}\Abs{B_{i}} & 0 & 0 &
 1-\tfrac12\sum_{i=1}^{n-1}\Abs{B_{i}}-\Abs{f_{1}^{N}(t_n)} \\[4pt]
 0 & f_{1}^{N}(t_n)^{\ast}A_{n-1} & 0 & 0 \\[4pt]
 0 & 0 & f_{1}^{N}(t_n)\,A_{n-1} & 0 \\[4pt]
 \tfrac12\sum_{i=1}^{n-1}\Abs{B_{i}} & 0 & 0 &
 \tfrac12\sum_{i=1}^{n-1}\Abs{B_{i}}+\Abs{f_{1}^{N}(t_n)}
\end{pmatrix}~,
\end{equation}
\end{widetext}
where the arguments of $B_i=B_i(t_{n;n-i})$ and $A_{n-1}=A_{n-1}(t_{n-1;1})$
have been suppressed for readability. Their explicit forms are given in
Ref.~\cite{ZammitMemory2026}; for the equal readout times considered below,
$A_{n-1}$ is real and positive. Comparing \cref{eq_mapn} with the
single-use map of \cref{eq_dyn_map_1}, the memory of the previous $n-1$ uses is
carried entirely by the two quantities $A_{n-1}$ and $B_{i}$. 

We specialise to a chain engineered for perfect state
transfer~\cite{Christandl2004}, for which a single use reaches unit
fidelity, $|f_{1}^{N}(\tstar)|=1$, at the optimal time $\tstar$; we work
in units in which $\tstar=\pi/2$. In particular, we take the $XX$ chain with
nearest-neighbour couplings $J_{i}=\sqrt{i(N-i)}$, $i=1,\dots,N-1$, for
which the end-to-end amplitude takes the form
$f_{1}^{N}(t)=(-i\sin t)^{N-1}$, and fix the chain length to $N=10$
throughout. Away from $\tstar$ the transfer is imperfect, and the
experimental imperfection considered here is an error in the readout
timing. We model a timer with a fixed
systematic offset, so that every use is read out at the same shifted time
\begin{equation}\label{eq_timing}
    t = (1-\varepsilon)\,\tstar~,
    \qquad \varepsilon\in[0,\varepsilon_{\max}]~,
\end{equation}
with $\varepsilon$ the fractional timing error, referred to below as the timing offset. Because the fidelity is
symmetric about $\tstar$ we consider offsets on one side, up to
$\varepsilon_{\max}=0.2$. Setting $t_1=\dots=t_n=t$ in \cref{eq_mapn}, the
$n^{\mathrm{th}}$-use channel reduces to a one-parameter family
\begin{equation}\label{eq_family}
    \Phin(t)\;\equiv\;\Phi_{n,\varepsilon}~,
    \qquad n\in\{1,\dots,M\},
\end{equation}
labelled by the use number $n$ and the timing offset $\varepsilon$. We take
$M=5$, giving the five channels---the five ``uses''---that our
classifier must tell apart. Throughout we index the classes $0,1,\dots,4$,
so that class $0$ is the first use ($n=1$) and class $4$ is the fifth use
($n=5$).

\subsection{$U(1)$ covariance and reduced coordinates}\label{sec_covariance}

Writing a qubit state through its Bloch vector $\bm{r}=(r_x,r_y,r_z)$ as $\hat{\rho}=\tfrac12(\mathds{1}+\bm{r}\cdot\hat{\bm{\sigma}})$,
the number-conservation property of \cref{eq_u1} makes every
$\Phi_{n,\varepsilon}$ covariant under rotations about the $z$
axis,
\begin{equation}\label{eq_covariance}
    \Phi_{n,\varepsilon}\!\left(e^{-i\theta \hat{\sigma}^{z}/2}\hat{\rho}\,e^{i\theta \hat{\sigma}^{z}/2}\right)
    = e^{-i\theta \hat{\sigma}^{z}/2}\,\Phi_{n,\varepsilon}(\hat{\rho})\,e^{i\theta \hat{\sigma}^{z}/2}
\end{equation}
for all $\theta$. On the Bloch ball this is a rotation of $(r_x,r_y)$
through $\theta$ at fixed $r_z$, so the action of the channel depends on
the input only through the pair
\begin{equation}\label{eq_reduced}
    (\rperp,\,r_z), \qquad \rperp=\sqrt{r_x^{2}+r_y^{2}}~,
\end{equation}
and returns an output pair $(\rperp',r_z')$ together with an azimuthal
phase that carries no information about $n$. The input azimuthal angle is
therefore a superfluous coordinate. This observation means that the discrimination
problem is intrinsically two-dimensional in $(\rperp,r_z)$ for each Bloch vector. This motivates 
the reduced input representation for the network.

The action of each use on the Bloch ball can be made explicit. For equal
readout times, $f_{1}^{N}(t)=(-i\sin t)^{N-1}=|f_{1}^{N}(t)|\,e^{i\gamma}$,
where the phase $\gamma=-(N-1)\pi/2$ (modulo $2\pi$) does not depend on $t$,
since $\sin t>0$ throughout the interval considered. The map of
\cref{eq_mapn} then acts on the Bloch vector as the affine transformation
\begin{equation}\label{eq_affine}
\begin{aligned}
    \begin{pmatrix} r_x' \\ r_y' \end{pmatrix} &=
    \big|f_{1}^{N}(t)\big|A_{n-1}
    \begin{pmatrix} \cos\gamma & -\sin\gamma \\ \sin\gamma & \cos\gamma \end{pmatrix}
    \begin{pmatrix} r_x \\ r_y \end{pmatrix}~, \\
    r_z' &= \big|f_{1}^{N}(t)\big|^{2}\,r_z
    + 1-\big|f_{1}^{N}(t)\big|^{2}-\sum_{i=1}^{n-1}\Abs{B_{i}}~,
\end{aligned}
\end{equation}
that is, a rotation through $\gamma$ about the $z$ axis, a contraction of
the transverse and longitudinal components, and a shift along $z$. For the
chain of length $N=10$ considered here, $\gamma=-\pi/2$. In the
reduced coordinates the first line becomes
$\rperp'=|f_{1}^{N}(t)|A_{n-1}\rperp$, so that $\rperp$ and $r_z$
transform independently of one another.

\begin{figure*}[t]
    \centering
    \includegraphics[width=\textwidth]{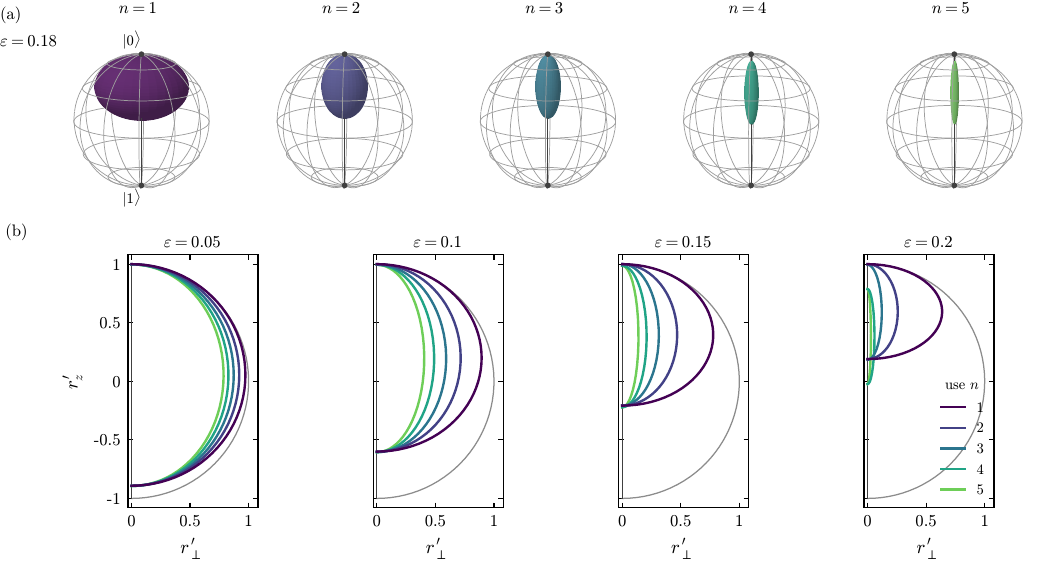}
    \caption{Action of the map $\Phi_{n,\varepsilon}$ on the Bloch ball for
    the first five uses of the chain. (a) Image of the Bloch ball under the map of each use at
    $\varepsilon=0.18$. Each image is a
    spheroid about the $z$ axis. The transverse semi-axis
    $|f_{1}^{N}(t)|A_{n-1}$ shrinks from use to use, the semi-axis
    $|f_{1}^{N}(t)|^{2}$ along $z$ is the same for every use, and the centre $z_{0}$ shifts downward.
    (b) Sections of the images in the reduced output coordinates
    $(\rperp',r_z')$ for $\varepsilon=0.05$, $0.1$, $0.15$ and $0.2$. The
    grey semicircle is the boundary of the Bloch ball.}
    \label{fig_bloch}
\end{figure*}

Each use therefore maps the Bloch ball onto a spheroid which is symmetric
about the $z$ axis, as shown in \cref{fig_bloch}. Its semi-axis along $z$,
$|f_{1}^{N}(t)|^{2}$, is the same for every use. The memory of the previous
uses thus enters only through the transverse semi-axis
$|f_{1}^{N}(t)|A_{n-1}$ and through the position of the centre,
$z_{0}=1-|f_{1}^{N}(t)|^{2}-\sum_{i=1}^{n-1}\Abs{B_{i}}$. For the chain of
length $N=10$, the former shrinks strictly from use to use for every
$\varepsilon>0$ in the interval. At $\varepsilon=0.18$, for example, it
decreases from $0.69$ for the first use to $0.06$ for the fifth. For the
same chain, the latter, relative to that of the first use, moves by at most
$0.006$ for the second and third uses. It moves appreciably only for the
fourth and fifth uses at large timing offsets, by up to $0.21$ at
$\varepsilon=0.2$. At $\varepsilon=0$ the five images
coincide with the whole Bloch ball, in agreement with
$\Phi_{n,0}=\Phi_{1,0}$. From \cref{fig_bloch}(b) it can also be seen that for $\varepsilon = 0.2$, the spheroids of different uses can overlap, so that an output state alone does not, in general, identify the use which produced it.

\section{Classification task and methods}\label{sec_methods}

Here we detail the task that was solved by the classifier. A single
labelled example consists of the Bloch vector $\bm{r}$ of the initial state of the sender 
$\hat{\rho}$, the Bloch vector $\bm{r}'$ of the state of the receiver $\Phi_{n,\varepsilon}(\hat{\rho})$, and the number of the use which produced them,
\begin{equation}\label{eq_example}
    \left(\bm{r},\,\bm{r}',\,n\right),
    \qquad n\in\{1,\dots,5\}~.
\end{equation}
The classifier observes only the pair of Bloch
vectors and must return the use $n$, encoded as the class $n-1$.

Two parameterisations of the input are considered. The first supplies
the raw components of both Bloch vectors, giving the six-dimensional
input $(r_{x}, r_{y}, r_{z}, r'_{x}, r'_{y}, r'_{z})$. The second
exploits the $U(1)$ covariance of the channel established in
\cref{sec_model}, and supplies instead the four-dimensional
input $(\rperp, r_{z}, \rperp', r'_{z})$. The former will be referred to as
the raw parameterisation and the latter as the reduced one. 

\subsection{Data generation}\label{sec_data}

Input states are pure states sampled from the Haar measure on the Bloch
sphere \cite{Mezzadri2007}. Fresh states are drawn for every combination
of use and timing, so that no input state is shared between classes.
Recalling that $M=5$ is the number of uses considered (\cref{sec_repeated}), a data set specified by $N_{s}$ states per use and timing, and
$N_{t}$ timings, contains
\begin{equation}\label{eq_size}
    N_{\mathrm{tot}} = M\, N_{s} N_{t}
\end{equation}
independently sampled examples, with the appropriate
$\Phi_{n,\varepsilon}$ applied to each block of $N_{s}$ states.

Training timings are taken as an equally spaced grid of $N_{t}$ points spanning
$[0,\varepsilon_{\max}]$, endpoints included, so that the grid contains the perfect
transfer time $\varepsilon=0$. At $\varepsilon=0$ the chain is emptied at 
every use and the five maps coincide, $\Phi_{n,0}=\Phi_{1,0}$ for all $n$, 
so that examples generated at this timing carry no information on the use 
number. We nonetheless retain this timing, since it is a physical 
operating point of the timer, and the one an experiment aims for.

Three test sets are used, all generated independently of the training
data. The first, termed the common test set, is used to score every
trained model. Its timings are constructed in two parts: the $99$
midpoints of the finest training grid ($N_{t}=100$), which lie as far from
its trained timings as the grid permits, and a further $50$ timings drawn
uniformly from the interval. With $100$ states per use and timing, the common 
test set contains $74{,}500$ examples. The second, termed the master test set,
is used for the headline evaluation of a single model per parameterisation.
It contains $10^{5}$ timings drawn uniformly from the interval, each paired
with a fresh state for every use, giving $10^{5}$ examples per use and
$5\times10^{5}$ in total. The third, termed the repeated test sets, consists of $100$ realisations of a test set, each using
the $99$ midpoint timings above with $100$ freshly sampled states per use
and timing. Uniformly drawn timings coincide with a trained timing with
probability zero, being draws from a continuous distribution. As such, every 
quoted test accuracy is evaluated on data absent from training.

\subsection{The classifier}\label{sec_network}

The classifier is a fully connected feed-forward neural network, that
is, a parameterised map $f_{\bm{\theta}}:\mathbb{R}^{d}\to\mathbb{R}^{5}$
built as a composition of affine transformations and element-wise
non-linearities. Each layer acts on the output of the previous one
according to
\begin{equation}\label{eq_layer}
    \bm{a}^{(\ell)} =
    g\!\left(W^{(\ell)}\bm{a}^{(\ell-1)}+\bm{b}^{(\ell)}\right)~,
\end{equation}
where $W^{(\ell)}$, $\bm{b}^{(\ell)}$, and $g$ are, respectively,
the weight matrix, the bias vector of the $\ell^{\mathrm{th}}$ layer, and 
the activation function. The final layer replaces $g$ with the softmax
\begin{equation}\label{eq_softmax}
    p_{c} = \frac{e^{z_{c}}}{\sum_{j=0}^{4}e^{z_{j}}}~,
\end{equation}
which converts the five outputs $z_{j}$ into a probability distribution
over the use labels; the predicted label is taken as
$\arg\max_{c} p_{c}$. Training consists of adjusting the parameters
$\bm{\theta}=\{W^{(\ell)},\bm{b}^{(\ell)}\}$ so as to minimise the
categorical cross-entropy loss
\begin{equation}\label{eq_loss}
    \mathcal{L} = -\sum_{c=0}^{4} y_{c}\ln p_{c}~,
\end{equation}
averaged over a batch, where $y_{c}$ is the one-hot encoding of the true
label. Minimisation proceeds iteratively over mini-batches of the
training data, so that each update is computed from a stochastic
estimate of the gradient, obtained through backpropagation. The updates
themselves are performed by the Adam optimiser \cite{kingmaAdam2015},
a first-order method which assigns to each parameter an adaptive step
size, determined from running estimates of the first and second moments
of its gradient (see, for example, \cite{Carleo2019, Dunjko2018} for
reviews of these methods in physical contexts). Each network is trained
for $1500$ epochs, and the weights retained are those of the epoch with the
highest accuracy on the validation set, i.e., the $20\%$ of the generated
data withheld from the gradient updates. The architecture and training configuration are summarised in
\cref{tab_arch}, and are held fixed across the entirety of this work.

\begin{table}[t]
\caption{Architecture and training configuration, common to every model
trained in this work. The two parameterisations differ only in the input
dimension, giving $3141$ and $3013$ trainable parameters for the raw and
reduced inputs, respectively.}
\label{tab_arch}
\begin{ruledtabular}
\begin{tabular}{ll}
Input dimension            & $6$ (raw) / $4$ (reduced) \\
Hidden layers              & Dense$(64)$, Dense$(32)$, Dense$(16)$ \\
Hidden activation          & ReLU \\
Output layer               & Dense$(5)$, softmax \\
Loss                       & categorical cross-entropy \\
Optimiser                  & Adam, learning rate $0.005$ \\
Epochs                     & $1500$ \\
Batch size                 & $300$ \\
Train/validation split     & $80/20$ \\
Retained weights           & best validation accuracy \\

\end{tabular}
\end{ruledtabular}
\end{table}

\subsection{Evaluation protocol and metrics}\label{sec_metrics}

\begin{figure}[t]
    \centering
    \includegraphics[width=\linewidth]{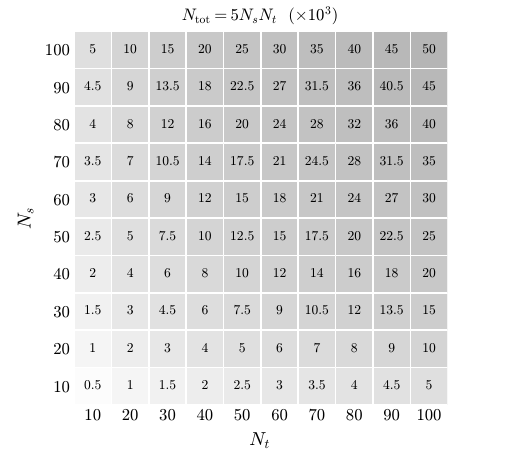}
    \caption{The training grid. Each cell corresponds to a data set with
    $N_{t}$ equally spaced timings and $N_{s}$ states per use and timing,
    and gives its size $N_{\mathrm{tot}}=5N_{s}N_{t}$ in thousands. Cells
    mirrored across the diagonal $N_{s}=N_{t}$ contain the same number of
    examples. For each point in the grid $100$ independent realisations of
    the data set are generated.}
    \label{fig_grid}
\end{figure}

Model quality is assessed along three axes. The first, termed training 
variability, asks what model accuracy should be expected from a training set 
of a given size. Given that $N_{\mathrm{tot}}$ depends on both $N_{s}$ and $N_{t}$,
we let both range from $10$ to $100$ in steps of $10$ and consider all data set
sizes in the grid shown in \cref{fig_grid}. For each point in the grid, $100$ data set realisations are generated, and one model is trained on each from
randomly initialised weights. Once trained, every model is
scored on the common test set. The raw and reduced models of a given realisation are
trained on the same data set, so that the two parameterisations can be
compared pairwise. The second, termed generalisation, asks how well a
single trained model performs on unseen states and timings. For each
parameterisation, a representative model is selected among the $100$
trained at $N_{s}=N_{t}=100$, namely the most typical one, whose accuracy
and loss on the common test set lie closest to the median values for that
data set size. These models are then evaluated on the master test set.
The third, termed test variability, asks how much the accuracy of a typical
model depends on the particular test set used to evaluate it. For this, the
same representative models are scored on $100$ realisations of a test set
of fixed size. 

The primary diagnostic is the row-normalised confusion matrix $C$, whose
entry $C_{ij}$ gives the fraction of class-$i$ examples assigned to class
$j$; the diagonal entries are the per-use accuracies. However, the use
labels are ordered, and a confusion matrix treats every error equally.
Mislabelling the third use as the first is a graver error than
mislabelling it as the second. To capture this, each row of $C$ is
additionally scored with the earth-mover (1-Wasserstein) distance
\cite{Rubner2000} to the ideal prediction, taken on the ordered label
set with ground metric $d(i,j)=|i-j|$. In one dimension this distance
is given by
\begin{equation}\label{eq_emd}
    W_{1}^{(i)} = \sum_{j}\left|i-j\right|C_{ij}~,
\end{equation}
i.e., the expected displacement, in units of the use number, of the
predictions for true class $i$; the class-averaged value
\begin{equation}
    \overline{W}_{1} = \frac{1}{5}\sum_{i}W^{(i)}_{1}
\end{equation}
condenses the error structure into a single figure and is directly interpretable as 
the average amount by which a prediction is incorrect, in units of uses.
A perfect classifier gives
$\overline{W}_{1}=0$, while uniformly random predictions give
$\overline{W}_{1}=1.6$ for five classes. A related quantity, the mean
displacement per error $D=\overline{W}_{1}/(1-\overline{A})$, where
\begin{equation}
    \overline{A} = \frac{1}{5}\sum_{i} C_{ii}
\end{equation}
is the overall accuracy, distinguishes structured from unstructured mistakes. Given that the prediction is wrong, $D$ is the average number of uses by which
the prediction is off. A value of $D$ near $1$ indicates a confusion between neighbouring uses only, while values approaching $2$, the value for uniformly random predictions, indicate errors scattered without regard to label order.

\section{Results}\label{sec_results}

\subsection{Classification accuracy and generalisation}\label{sec_acc}

\begin{figure*}[t]
    \centering
    \includegraphics[width=\textwidth]{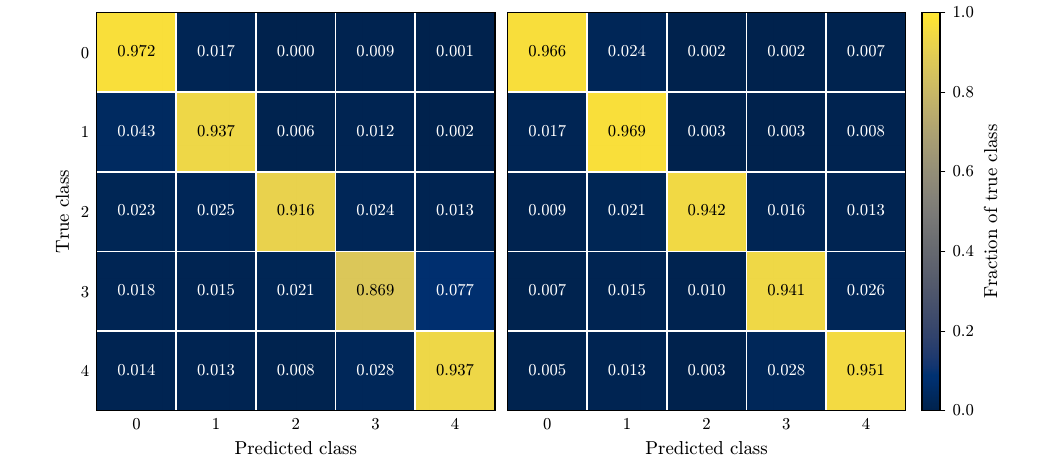}
    \caption{Row-normalised confusion matrices of the representative models
    for the raw (left) and reduced (right) parameterisations, evaluated on
    the master test set. Entry $(i,j)$ is the fraction of examples of true
    class $i$ assigned to class $j$, and class $c$ corresponds to the
    $(c+1)^{\mathrm{th}}$ use of the chain. The test set contains $10^{5}$
    examples per class, each with a fresh Haar-random input state and a
    timing offset drawn uniformly from $\varepsilon\in[0,0.2]$. Both
    models are the most typical of the $100$ trained at
    $N_{s}=N_{t}=100$.}
    \label{fig_conf_master}
\end{figure*}

The performance of the classifier is first assessed on the master test
set, using the representative model of each parameterisation described in
\cref{sec_metrics}. Every example in this set carries a fresh input state
and a timing offset drawn uniformly from $\varepsilon\in[0,0.2]$.
The row-normalised confusion
matrices are shown in \cref{fig_conf_master}. The overall accuracy is
$0.926$ for the raw parameterisation and $0.954$ for the reduced one.

The per-use accuracies, read off the diagonals, range from $0.869$ to
$0.972$ for the raw parameterisation and from $0.941$ to $0.969$ for the
reduced one. The reduced parameterisation improves the accuracy for the
second to the fifth use, while for the first use the two are comparable
($0.972$ and $0.966$). The largest gain occurs for the fourth use (class
$3$), whose accuracy rises from $0.869$ to $0.941$. In particular, the raw
model assigns $7.7\%$ of fourth-use examples to the fifth use, while only
$2.8\%$ of fifth-use examples are assigned to the fourth. The reduced model
removes this asymmetry, the two fractions becoming $2.6\%$ and $2.8\%$. A
second difference can be seen in the first column. The raw model assigns between
$1.4\%$ and $4.3\%$ of the examples of every later use to the first use,
whereas the reduced model assigns between $0.5\%$ and $1.7\%$.

Physically, these differences are consistent with the $U(1)$ covariance of
the maps. The azimuthal angle of the input carries no
information on the use number, so the raw network must learn to disregard
it from the training data alone. The reduced parameterisation removes this
superfluous coordinate before training, leaving the network to model only the
two-dimensional dependence on $(\rperp, r_{z})$. 

These accuracies are consistent with those of the full set of models
trained at the same grid point. On the common test set, the $100$ models
trained at $N_{s}=N_{t}=100$ reach mean accuracies of $0.9223\pm0.0035$ and
$0.9508\pm0.0023$ for the raw and reduced parameterisations, respectively,
the uncertainty being the standard deviation across training realisations.

\begin{figure*}[t]
    \centering
    \includegraphics[width=\textwidth]{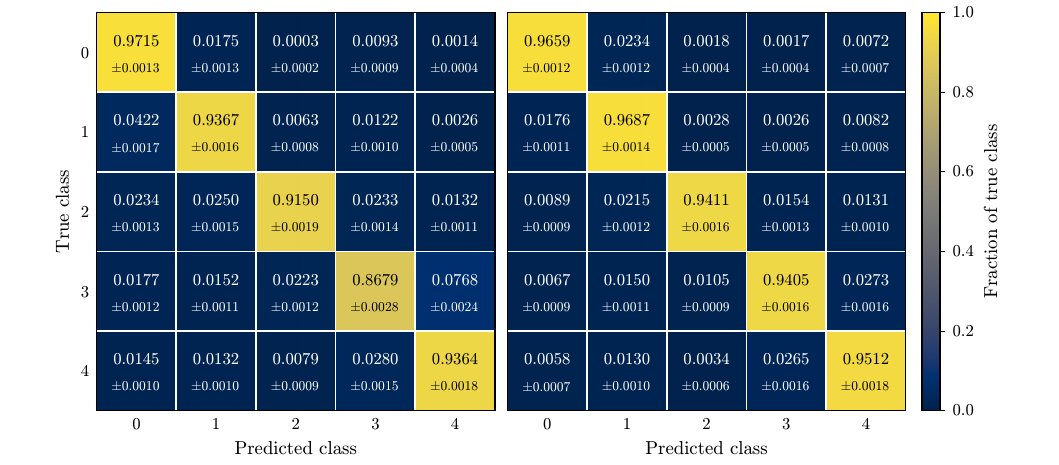}
    \caption{Row-normalised confusion matrices of the same representative
    models as in \cref{fig_conf_master}, for the raw (left) and reduced
    (right) parameterisations, averaged over $100$ repeated test sets. Each
    set contains $100$ freshly sampled states per use at each of the $99$
    midpoints of the $N_{t}=100$ training grid, i.e., $9{,}900$ examples per
    class. The upper number in each cell is the mean over the test sets and
    the lower number is the standard deviation across them.}
    \label{fig_conf_avg}
\end{figure*}

The averaged confusion matrices obtained from scoring the models on the $100$ 
test set realisations are shown in \cref{fig_conf_avg}. The accuracy
varies across the test sets by only $\pm0.0009$ (raw) and $\pm0.0008$ (reduced)
about means of $0.9255$ and $0.9535$, and no individual entry varies by more 
than $0.003$. These spreads are three to four times smaller than the spread across
training realisations quoted above. The dominant uncertainty in the reported
accuracies is therefore the training of the network, not the sampling of
the test data. Moreover, the averaged matrices agree with those of
\cref{fig_conf_master} to within $0.0012$ in every entry. This agreement is
informative because the two test sets probe different timings. The midpoint
timings lie as far as the training grid permits from any trained timing,
whereas the timings of the master set are spread uniformly over the
interval. Equal accuracies on the two indicate that interpolating between
trained timings costs the classifier no measurable accuracy.

\subsection{Data set size and efficiency}\label{sec_size}

\begin{figure*}[t]
    \centering
    \includegraphics[width=\textwidth]{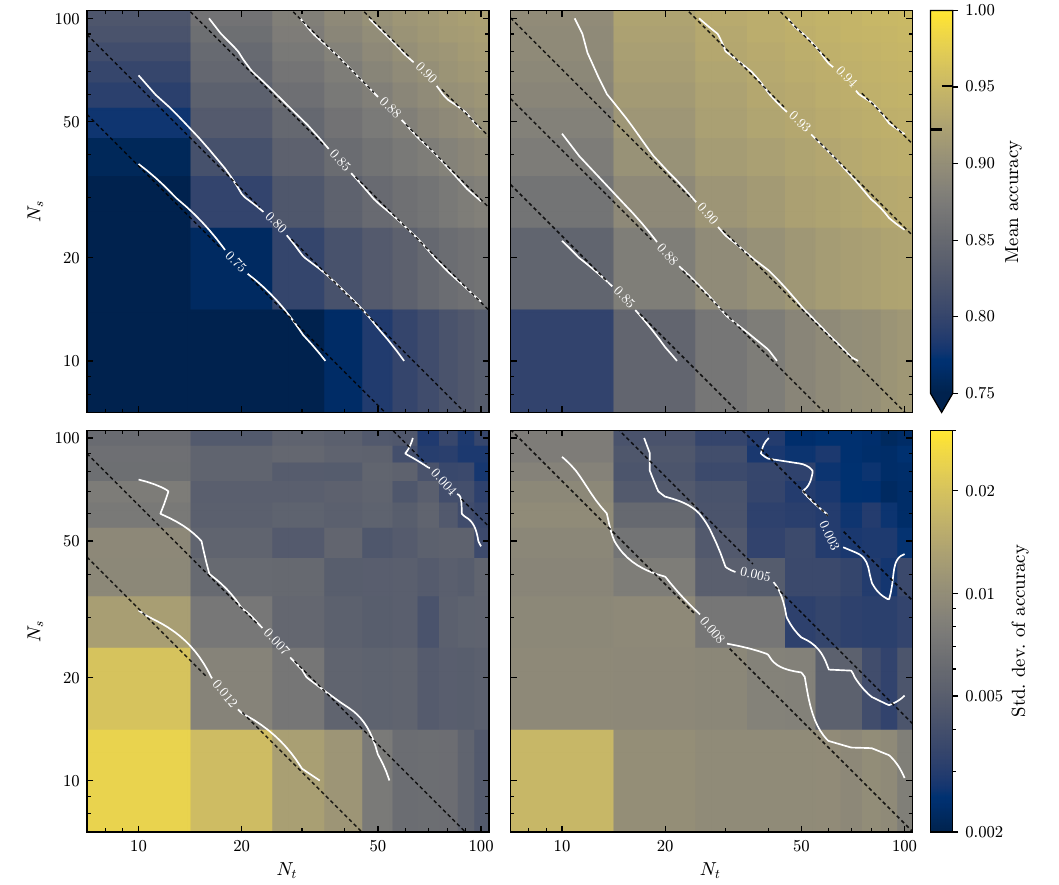}
    \caption{Mean (top) and standard deviation (bottom) of the test accuracy
    over $100$ training realisations at each point of the training grid, for
    the raw (left) and reduced (right) parameterisations. Solid white curves
    are contours of the measured means (top) and standard deviations
    (bottom), obtained by bilinear interpolation in $(\ln N_{t},\ln N_{s})$.
    Each dashed line marks constant $N_{s}N_{t}$, i.e., constant data set
    size, and is included as a guideline, since a quantity depending on $N_{s}N_{t}$
    alone would follow it. The two short black marks inside the colour bar of the means
    give the maxima of the left and right panels, respectively. The standard deviation
    is shown on a logarithmic colour scale.}
    \label{fig_size}
\end{figure*}

The dependence of the classifier on the amount of training data is shown in
\cref{fig_size}. The upper panels give the mean test accuracy of the $100$
models trained at each point of the $(N_{t},N_{s})$ grid, evaluated on the
common test set. The lower panels give the standard deviation of the same
accuracies across the $100$ training realisations. Both axes are
logarithmic, and the dashed lines mark constant $N_{s}N_{t}$, i.e., a
constant data set size $N_{\mathrm{tot}}$. The mean
accuracy rises monotonically with both $N_{s}$ and $N_{t}$. For the raw
parameterisation it rises from $0.515$ at $N_{s}=N_{t}=10$ to $0.922$ at
$N_{s}=N_{t}=100$, and for the reduced one from $0.797$ to $0.951$. The
increase slows towards the largest data sets, although neither
parameterisation has fully saturated within the grid.

The iso-accuracy contours in \cref{fig_size} can be seen to follow the dashed lines of
constant $N_{s}N_{t}$ across the grid, apart from the $N_{t}=10$ edge discussed below, for both parameterisations.
If either $N_{s}$ or $N_{t}$ had a larger effect on the obtained accuracy, this would
be seen as a deviation between the iso-accuracy curves and the dashed lines
in \cref{fig_size}. As these lines essentially always overlap, we conclude that
the accuracy depends essentially only on the
product $N_{s}N_{t}$, and not on how the data budget is divided between
states and timings, i.e., the number of sampled states and the
number of sampled timings are equally important. The only systematic dependence on
the division of the data is a lower mean accuracy along the edge
$N_{t}=10$, by up to $0.014$ relative to the mirrored cell, possibly in
part because one in ten training timings is then $\varepsilon=0$.

This observation also bears on the generalisation in time discussed in
\cref{sec_acc}. At fixed $N_{s}N_{t}$, reducing $N_{t}$ from $100$ to $10$
widens the largest gap in $\varepsilon$ between a test timing and the
nearest trained timing by a factor of eleven, from $0.001$ to $0.011$.
That this leaves the accuracy essentially unchanged indicates that the
network interpolates smoothly in the timing offset, rather than relying on
having seen timings close to those it is tested on.

Since the accuracy depends essentially on $N_{s}N_{t}$ alone, the data set
size needed to reach a given accuracy can be stated as a single number.
\Cref{tab_size} lists, for several target accuracies, the smallest
$N_{\mathrm{tot}}$ from which every measured combination of $N_{s}$ and
$N_{t}$ reaches the target on average. The reduced parameterisation reaches
every overlapping target with between four and seven times fewer examples 
than the raw one. In particular, a mean accuracy of $0.90$ requires
$N_{\mathrm{tot}}=24{,}000$ examples with the raw input, but only $6{,}000$
with the reduced input. The raw parameterisation does not reach $0.93$
within the grid at all.

\begin{table}[t]
\caption{Smallest data set size $N_{\mathrm{tot}}=5N_{s}N_{t}$ from which
every measured combination of $N_{s}$ and $N_{t}$ reaches the target mean
accuracy, for both parameterisations. Sizes include the $20\%$ of the data
withheld for validation. A dash indicates a target not reached within the
grid, whose largest data set has $N_{\mathrm{tot}}=50{,}000$.}
\label{tab_size}
\begin{ruledtabular}
\begin{tabular}{ccc}
Target accuracy & Raw & Reduced \\
\hline
$0.85$ & $8{,}000$  & $1{,}500$  \\
$0.88$ & $15{,}000$ & $2{,}500$  \\
$0.90$ & $24{,}000$ & $6{,}000$  \\
$0.92$ & $50{,}000$ & $7{,}500$  \\
$0.93$ & --         & $12{,}500$ \\
$0.94$ & --         & $24{,}000$ \\
$0.95$ & --         & $50{,}000$ \\
\end{tabular}
\end{ruledtabular}
\end{table}

The advantage of the reduced parameterisation holds throughout the grid.
Compared realisation by realisation, the reduced
model is the more accurate one in every one of the $10{,}000$ pairs. The
mean gap is largest for the smallest data sets, where it reaches $0.28$ at
$N_{s}=N_{t}=10$, and shrinks to $0.0285$ at $N_{s}=N_{t}=100$. Even the smallest
difference within a single pair is $0.011$. A clue to
the origin of this gap is provided by a supplementary set of trainings at
$(N_{t},N_{s})=(10,50)$, comprising ten data sets each trained from ten
independent initialisations. At the final epoch, the raw models reach a
mean training accuracy of $0.90$ but a validation accuracy of only $0.74$,
whereas the reduced models reach $0.87$ and $0.83$, respectively. The raw
networks therefore over-fit markedly when data is scarce, while the reduced
ones do not. This is consistent with the raw networks having to learn
the irrelevance of the azimuthal angle from a limited number of examples.

The lower panels of \cref{fig_size} 
show how far a single trained network may stray from these averages. The
standard deviation across realisations decreases with the amount of data,
from $0.024$ (raw) and $0.017$ (reduced) at $N_{s}=N_{t}=10$ to $0.0035$ 
and $0.0023$ at $N_{s}=N_{t}=100$. Like the mean, the spread depends 
essentially on the data set size alone: cells sharing the same $N_{s}N_{t}$
differ systematically by no more than about $5\%$, comparable with the 
sampling uncertainty of about $7\%$. This sampling noise is also why the 
contours of the lower panels follow their dashed lines less
closely than those of the upper panels.

\subsection{Severity of misclassifications}\label{sec_emd}

The severity of the residual errors is quantified with the earth-mover
distance of \cref{eq_emd}, in units of the use number. On the master test
set, the class-averaged distance is $\overline{W}_{1}=0.114$ for the raw
parameterisation and $0.074$ for the reduced one. The
reduced parameterisation thus lowers $\overline{W}_{1}$ by about a third.
The repeated test sets give $0.116\pm0.001$ and $0.075\pm0.001$,
respectively, in agreement with the master set.

The mean displacement given that the prediction is
incorrect, $D$, equals $1.55$ and $1.60$ for the raw and  the reduced 
parameterisations, respectively. Both values lie between $1$, the value for
errors confined to neighbouring uses, and $2$, the value for errors
scattered uniformly over the other uses. In particular, about two thirds of
the misclassified examples ($65\%$ and $63\%$, respectively) are assigned to
a neighbouring use. The remaining third are displaced by two or more uses,
as seen in the far corners of \cref{fig_conf_master}. For example, the raw
model assigns $1.4\%$ of fifth-use examples to the first use. Taken together,
these values suggest that the classifier identifies the use number correctly
in over $90\%$ of cases, and that when misclassifications do occur, they
most likely fall within one use of the truth.

\begin{figure}[t]
    \centering
    \includegraphics[width=\linewidth]{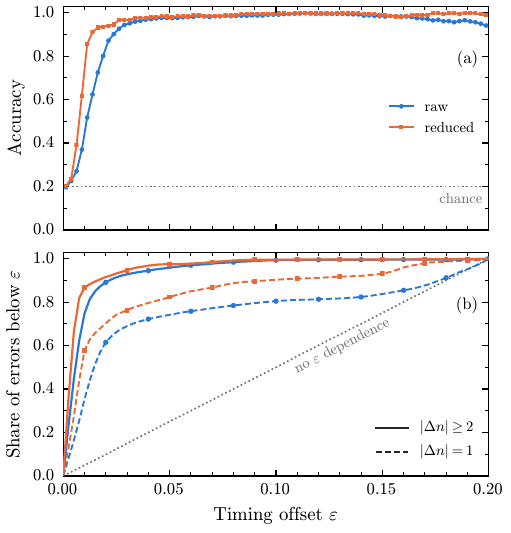}
    \caption{Errors of the representative models on the master test set,
    resolved by the timing offset $\varepsilon$ of the examples, for the
    raw and reduced parameterisations. The examples are grouped into $80$
    bins of width $0.0025$ in $\varepsilon$, each containing about
    $6{,}250$ examples with equal numbers per use. (a)~Fraction of correctly
    classified examples in each bin; the standard error of each point is
    below $0.007$, and the dotted line marks the chance level, $1/5$.
    (b)~Cumulative share of the misclassified examples whose offset lies
    below $\varepsilon$, for errors displaced by two or more uses (solid)
    and by one use (dashed). The dotted line is what an error curve would follow if the error rate did not depend on~$\varepsilon$.}
    \label{fig_acc_eps}
\end{figure}

The far-displaced errors originate predominantly from timing offsets close to
$\varepsilon=0$, as shown in \cref{fig_acc_eps}, where the errors on the
master test set are resolved by the timing offset of each example. For
noiseless Bloch vectors, the five maps coincide exactly only at this point,
so the classes near it remain distinguishable in principle, but their
differences become small compared with the resolution of the network. Both
the master test set and the training grid sample this region. A classifier
unable to resolve such small differences can do no better than a guess that
is only weakly tied to the true use, and errors of this kind need not fall
on a neighbouring use. Indeed, \cref{fig_acc_eps}(a) shows both models at
chance level for $\varepsilon<0.0025$, with the accuracy reaching $0.9$
only by $\varepsilon=0.015$ for the reduced parameterisation and by
$\varepsilon=0.025$ for the raw one. Above $\varepsilon=0.02$, the
accuracy is $0.976$ (raw) and $0.989$ (reduced). As \cref{fig_acc_eps}(b)
shows, offsets below $\varepsilon=0.02$, a tenth of the master test set,
account for $89\%$ (raw) and $92\%$ (reduced) of the errors displaced by
two or more uses, against $62\%$ and $70\%$ of those on a neighbouring
use.

\section{Conclusion}\label{sec_conclusion}
When a spin chain is used repeatedly for QST without being reset, the
excitations left behind by earlier transfers endow the channel with a
memory, so that the map governing each use depends on the number of uses
which preceded it. This work set out to determine whether that number can
be recovered from the data of a single transmission. The results show that
it can. For an $XX$ chain of length $N=10$ engineered for PST, and read out
with a fixed systematic timing offset of up to $20\%$, a small feed-forward
neural network given only the sender and receiver Bloch vectors identifies
which of the first five uses produced them. On a test set of fresh input
states with timing offsets drawn uniformly over this range, the
representative networks reach accuracies of $92.6\%$ with the raw Bloch
vectors and $95.4\%$ with the reduced coordinates.

Since every map is covariant under rotations about the $z$ axis,
the azimuthal angle of the input carries no information on the use number,
and the problem is intrinsically two-dimensional in $(\rperp,r_{z})$ for each Bloch vector.
Removing this superfluous coordinate before training made the network more
accurate in every one of the $10{,}000$ paired trainings, and reduced the
data needed for a given accuracy by a factor of between four and seven. The
gain was largest when data was scarce, where the raw networks were seen to over-fit, while the reduced ones did not.

Across the grid of data sets considered, the mean accuracy was found to depend on the
number of sampled states $N_{s}$ and of sampled timings $N_{t}$ essentially only through their 
product. The two are therefore equally important, and the data required for a target 
accuracy can be quoted as a single number, as done in \cref{tab_size}. Furthermore, since
high accuracies can be achieved even when widening the gap between trained and tested 
timings, the network appears to interpolate smoothly in the timing offset. In practice,
this suggests that the timing offset need not be sampled finely, provided that
the total number of examples is maintained.

The structure of the residual errors was quantified with an earth-mover
distance on the confusion matrix. The reduced coordinates lower the
class-averaged distance by about a third, from $0.114$ to $0.074$ uses.
About two thirds of the misclassified examples are assigned to a
neighbouring use, while the remainder are displaced by two or more uses.
The use number is therefore recovered as an ordered quantity only in part.
The far-displaced errors arise predominantly from timing offsets close to
$\varepsilon=0$: offsets below $0.02$, a tenth of the test examples,
account for about $90\%$ of them. The classifier relies on the same
memory which degrades the fidelity of later uses \cite{ZammitMemory2026},
and near the perfect transfer time there is little memory to read. Being
largely confined to this region, these errors are also the least consequential
ones, since the channels being confused there differ only slightly.

The results indicate that the memory of a quantum channel, ordinarily 
regarded as a source of error, also leaves a record of the channel's 
history in its input-output behaviour. This record can be read from transmission data alone, offering a route to the calibration and resource accounting of a quantum bus, without any additional probing of the medium.

\begin{acknowledgments}
The authors acknowledge funding from Grant Agreement No.~DTP-2024-13, Project ATTESTER financed by the Xjenza Malta, and the Malta Digital Innovation Authority, through the `R\&I Thematic Programmes: Digital Technologies Programme'. HZ also acknowledges funding by the Tertiary Education Scholarships Scheme (TESS).
\end{acknowledgments}

\section*{Data availability}
The data that support the findings of this article are openly available~\cite{zammit_2026_23337649}.
The raw training data sets and trained networks are available from the authors upon reasonable request.

\bibliography{n_use}

\end{document}